\documentstyle[preprint,epsfig,times,aps]{revtex}

\def\eg{{\it e.g., }}
\def\ie{{\it i.e., }}
\def\Z{{\bf Z}}
\def\R{{\bf R}}
\def\a{\alpha}
\def\s{\sigma}
\def\S{\Sigma}
\def\J{{\cal J}}
\def\d{\delta}
\def\I{{\cal I}}
\def\F{{\cal F}}
\def\M{{\cal M}}
\def\G{\Gamma}  
\def\P{\Phi}
\def\L{\Lambda}
\def\C{{\cal C}}
\date{}

\begin{document}

\draft
\tightenlines

\title{Zero-Temperature Dynamics of $\pm\,J$ Spin Glasses and Related
Models}
\author{A.~Gandolfi}
\address{Dipartimento di Matematica, Universit\`a di Roma Tor Vergata,
00133 Roma, ITALIA}
\author{C.M.~Newman}
\address{Courant Institute of Mathematical Sciences,
New York University, New York, NY 10012}
\author{D.L.~Stein}
\address{Departments of Physics and Mathematics, University of Arizona,
Tucson, AZ 85721}

\maketitle

\begin{abstract}
We study zero-temperature, stochastic Ising models $\sigma^t$ on $\Z^d$
with (disordered) nearest-neighbor couplings independently chosen from a
distribution $\mu$ on $\R$ and an initial spin configuration chosen
uniformly at random.  Given $d$, call $\mu$ type $\I$ (resp., type $\F$)
if, for {\it every\/} $x$ in $\Z^d$, $\sigma_x^t$ flips infinitely (resp.,
only finitely) many times as $t \to \infty$ (with probability one) --- or
else mixed type $\M$.  Models of type $\I$ and $\M$ exhibit a
zero-temperature version of ``local non-equilibration''.  For $d=1$, all
types occur and the type of any $\mu$ is easy to determine.  The main
result of this paper is a proof that for $d=2$, $\pm J$ models (where $\mu
= \a \delta_J + (1-\a)\delta_{-J}$) are type $\M$, unlike homogeneous
models (type $\I$) or continuous (finite mean) $\mu$'s (type $\F$). We also
prove that all other noncontinuous disordered systems are type $\M$ for any
$d\ge 2$.  The $\pm J$ proof is noteworthy in that it is much less ``local''
than the other (simpler) proof.
Homogeneous and $\pm J$ models for $d\ge 3$ remain an open
problem.
\end{abstract}

\bigskip


\section{Introduction and Results}
\label{sec:intro}

In this paper, we study a specific class of continuous time Markov processes
$\sigma^t = \sigma^t(\omega)$ with random environments. These correspond to
the zero-temperature stochastic dynamics of disordered nearest-neighbor Ising
models \cite{NNS,NS99,Jainold,Jainnew}. (Zero-temperature dynamics with
a different sort of disorder are studied in \cite{FIN}.) 
The state space is ${\cal S} = \{-1,+1\}^{\Z^d}$ and the initial
state $\s^0$ is a realization of i.i.d.~symmetric Bernoulli variables.
The only transitions are single spin flips, where $\s_x^{t+0} = -\s_x^{t-0}$,
and the transition rates depend on a realization $\J$ of i.i.d. random 
couplings $J_{x,y}\,$, indexed by nearest-neighbor pairs (with Euclidean distance
$||x-y|| = 1$) of sites in $\Z^d$, with common distribution $\mu$ on $\R$.
For a given $\J$, the rate for a flip at $x$ from state $\s^{t-0} = \s$
is $1$ or $1/2$ or $0$ according to whether 
\begin{equation}
\label{eq:one}
\Delta H_x (\sigma) \equiv 2 \sum_{y:\, ||y-x||=1}J_{x,y} \sigma_x \sigma_y\ 
\end{equation}
(corresponding to the change in energy) is negative or zero or positive. 
The joint distribution of $\J,\,\s^0$, and $\omega$ will be denoted $P$. 

Zero-temperature dynamics without disorder have been much studied in the
physics literature as a model of ``coarsening'' \cite{Bray} and more
recently because of the interesting phenomenon of persistence
\cite{St,De,DHP,MH,MS}.  A natural question in both the disordered and
non-disordered models is whether $\s^t$ has a limit (with $P$-probability
one) as $t \to \infty$ or equivalently whether for every $x$, $\s_x^t$
flips only finitely many times. More generally, one may call such an $x$ an
$\F$-site ($\F$ for finite) and otherwise an $\I$-site ($\I$ for infinite).
The nonexistence of a limit corresponds to the type of ``recurrence''
studied in a general context and applied to various interacting particle
systems in \cite{CK}.

The issue of whether $\s^t$ has no limit is the zero-temperature version of
whether there is ``local non-equilibration'' at positive temperature
\cite{NS99b}.  At {\it positive\/} temperature, local non-equilibration
concerns not the recurrence of the spin {\it configuration\/} $\sigma^t$
but rather of a dynamical probability measure $\nu_{t,\tau(t)}$
corresponding to averaging over the dynamics for times between $t-\tau(t)$
and $t$ (for fixed $\J, \sigma^0$ and dynamics realization $\omega$ up to
time $t-\tau(t)$).  For $\tau(t)$ growing slowly with $t$,
$\nu_{t,\tau(t)}$ should be (approximately) a pure Gibbs state at the given
temperature (on a lengthscale growing with $t$).  Local non-equilibration
would mean that the system does {\it not\/} converge to a single limiting
pure state as $t \to \infty$ (depending on $\J, \sigma^0,
\omega$). Although this type of non-equilibration has been proved to occur
for the $d=2$ homogeneous Ising model \cite{NS99b}, it is an open problem
whether it occurs at positive temperature for spin glasses (for any
$d\ge2$).  The focus of this paper is the study of the analogous problem at
zero temperature for certain classes of spin glasses and related disordered
systems.

By translation-ergodicity, the collection of $\F$-sites (resp., $\I$-sites)
has (with P-probability one) a well-defined non-random spatial density
$\rho_\F$ (resp., $\rho_\I$).  The densities $\rho_\F$ and $\rho_\I$ depend
only on $d$ and $\mu$ and of course satisfy $\rho_\F + \rho_\I = 1$. For
each $d$, one may then characterize $\mu$ (or more accurately, one should
characterize the pair $(d,\mu)$) as being type $\F$ or $\I$ or $\M$ (for
mixed) according to whether $\rho_\F(d,\mu) = 1$ or $\rho_\I(d,\mu) = 1$ or
$0<\rho_\F,\rho_\I<1$.

Before reviewing previous characterization results and presenting new ones,
we briefly discuss some important special cases of $\mu$.
Ferromagnetic models are those where $\mu$ is supported on $[0,\infty)$
(so that each $J_{x,y} \ge 0$) and homogeneous ones are those without
disorder (\ie, where $\mu = \d_{J'}$). In the homogeneous ferromagnet,
sites flip at rate $1$ or $1/2$ or $0$ according to whether they disagree
with a strict majority or exactly one half or a strict minority of their
nearest neighbors. Antiferromagnetic models are those with $J_{x,y} \le 0$;
on the lattice $\Z^d$, these are equivalent to ferromagnetic models under the
relabelling (or ``gauge'') transformation in which $\s_x \to -\s_x$ for each
$x$ on the odd sublattice while $J_{x,y} \to -J_{x,y}$ for every $\{x,y\}$
(leaving (\ref{eq:one}) unchanged). Spin glasses (of the Edwards-Anderson
type \cite{EA}) may be defined
as those models where $\mu$ is symmetric (under $J_{x,y} \to -J_{x,y}$) ---
the most popular examples being mean zero Gaussian distributions and the
$\pm J$ spin glasses, where $\mu = (1/2)\d_J + (1/2)\d_{-J}$ with $J>0$
(a standard review is \cite{BY}).
As we shall see, the family of measures $\mu = \a\d_J + (1-\a)\d_{-J}$ with
$\a \in [0,1]$, including homogeneous ferromagnets and antiferromagnets
and $\pm J$ spin glasses, is the most dificult to characterize. Henceforth,
we use the term $\pm J$ model to refer to any $\mu$ of the form
$\a\d_J + (1-\a)\d_{-J}$ with $J>0$ and $0 < \a < 1$. 

Our review of known results begins with a proposition classifying all
$\mu$'s for $d=1$. The type $\I$ nature of one-dimensional homogeneous
ferromagnets was stated in \cite{NNS}, but is equivalent to a result in
\cite{Arr} (see also \cite{CG}) because for $d=1$, the dynamics is the same
as that for annihilating random walks or the usual voter model.  It is
possible that other parts of the proposition may also not be new.

\medskip

{\bf Proposition 1.\/}  Set $d=1$. Then $\mu$ is type $\I$ if it is 
$\a\d_J + (1-\a)\d_{-J}$ with $J\ge0$ and $\a \in [0,1]$; $\mu$ is type 
$\F$ if it is either continuous or else of the form $\a\d_J + \beta \d_{-J} +\nu$
with $J>0$, $0 < \a + \beta < 1$ and a continuous $\nu$ supported on
$[-J,J]$; all other $\mu$'s are type $\M$. 

\medskip

{\bf Proof.\/}  Since $d=1$, any $\J = (J_{x,x+1} :\,x \in \Z)$ is equivalent
(by an appropriate gauge transformation) to a ferromagnetic model with
$J_{x,x+1}$ replaced by $|J_{x,x+1}|$. Hence, for the remainder of this proof,
we can and will assume that $\mu$ is replaced by $\overline\mu$, the common 
distribution of the $|J_{x,x+1}|$'s, and thus that each $J_{x,x+1}>0$.

If $\overline\mu = \d_J$, then it is trivially type $\I$ for $J=0$, while
for $J>0$, we have a homogeneous ferromagnet, for which a proof that it is type $\I$ 
may be found in \cite{NNS}. For any other $\overline\mu$, one looks for
sites $z$ such that
\begin{equation}
\label{eq:two}
J_{z,z+1} > \, J_{z-1,z}\, ,\,J_{z+1,z+2}\,.
\end{equation}
Since $\overline\mu \neq \d_J$, this has a strictly positive probability,
and hence, by translation-ergodicity, there will be (with $P$-probability one)
a doubly infinite sequence of such sites $z_n$ (with positive density).
The conditions (\ref{eq:two}) imply that 
$\Delta H_z(\s^t)$ and $\Delta H_{z+1}(\s^t)$ (see (\ref{eq:one})) are  
both negative or both positive according to whether $\s_z^t \s_{z+1}^t = -1$ or $+1$.
It follows that if $\s_z^0 \s_{z+1}^0 = +1$, then $\s_z^t$ and $\s_{z+1}^t$
will never flip, while $\s_z^0 \s_{z+1}^0 = -1$ implies that (with probability
one) one of them will flip exactly once and there will be no other flips
of either. This already shows that $\rho_\F>0$. 

If $\overline\mu$ is continuous, we may rely on the proof in \cite{NNS} or
argue as follows.  Restricting attention to an interval
$\{z,z+1,\dots,z'\}$ with $z=z_{n-1} +1$ and $z' = z_n$ (where $z_{n-1}$
and $z_n$ are successive sites from the special sequence defined above) and
times after $\s_z^t$ and $\s_{z'}^t$ have ceased flipping, we have a Markov
process with a finite state space --- the configurations of $(\s_x:\, z+1
\le x \le z'-1)$. Because of the continuity of $\overline\mu$, each flip in
this interval will {\it strictly\/} lower the energy,
\begin{equation}
\label{eq:three}
-\sum_{x=z}^{z'-1} J_{x,x+1}\s_x\s_{x+1}\,.
\end{equation} 
Since this energy is (for a fixed $\J$) bounded below, the process
in the interval must eventually stop flipping
and reach an absorbing state. Applying this argument to
every such interval, we conclude that a continuous $\overline\mu$ is type $\F$. 

If $\overline\mu = {\overline \a}\d_J + {\overline \nu}$ with $J>0,\,
0<{\overline \a}<1$ and ${\overline \nu}$ a continuous measure on
$[0,J]$, then we modify the above argument as follows. Instead of looking
for sites $z$ satisfying (\ref{eq:two}), we look for runs of the value $J$,
\ie for sites $z<w$ where
\begin{equation}
\label{eq:four}
J_{z-1,z}<J,\ J_{z,z+1}=J,\ J_{z+1,z+2}=J,\ \dots\ ,\ J_{w-1,w}=J,\
J_{w,w+1}<J\,.
\end{equation}
Now let $\{z_n,z_n +1,\dots,w_n\}$, as $n$ varies over $\Z$, be the doubly
infinite sequence of run intervals.  Focusing on the configurations in one
of these intervals, and noting that the transition rates in that interval
do {\it not\/} depend on the values of $\s_{z_n-1}$ or $\s_{w_n+1}$, we
observe that the two constant configurations are absorbing and accessible
from any other configuration, so that the process eventually reaches one of
these two absorbing configurations. To conclude that this $\overline\mu$ is
type $\F$, we need to show for each $n$, that (after $\s_{w_{n-1}}^t$ and
$\s_{z_n}^t$ have ceased flipping) the configuration in the interval
$\{w_{n-1}, w_{n-1}+1,\dots,z_n\}$ will also reach an absorbing state. But
this follows exactly as in the argument above for continuous
$\overline\mu$, with the continuity of ${\overline \nu}$ replacing that of
$\overline\mu$.

To complete the proof of the proposition, it remains to show that if
$\overline\mu$ is neither $\d_J$ nor continuous nor of the form 
${\overline \a}\d_J + {\overline \nu}$ as above, then $\rho_\I >0$ ---
\ie some spins flip infinitely often. But for any $\overline\mu$ now
under consideration, there will exist some $J' >0$ and sites
$z'$ and $z''=z'+3$ such that $z'$ and $z''$ each satisfy (\ref{eq:two}),
\begin{equation}
\label{eq:five}
J_{z'+1,z'+2}=J_{z'+2,z'+3}\,(\equiv J_{z''-1,z''})\,=J'\, ,
\end{equation}
and
\begin{equation}
\label{eq:six}
\s_{z'}^0=\s_{z'+1}^0=+1\, ,\ \ \s_{z''}^0=\s_{z''+1}^0=-1\,.
\end{equation}
Under these circumstances, $\s_{z'+1}^t$ and $\s_{z'+3}^t$ ($\equiv\s_{z''}^t$)
will never flip, but $\s_{z'+2}^t$ will flip infinitely many times because
its flip rate will always be $1/2$. 

\medskip

Among the main results of \cite{NNS} are extensions of the conclusions of
Proposition 1 to $d=2$ for the homogeneous ferromagnet (or antiferromagnet) and
to $d\ge2$ for continuous $\mu$ (satisfying some conditions).
In particular,
it is proved there that a continuous $\mu$ with finite mean (\ie with
$E(|J_{x,y}|) < \infty$) is type $\F$ for any $d$.
(Certain continuous
$\mu$'s with infinite means are also shown in \cite{NNS} to be type $\F$ 
by the very different percolation-theoretic methods of \cite{NN}.)
The continuous finite mean $\mu$ result is actually a corollary of the following
more general theorem about flips that strictly decrease
the energy, which we will apply to $\pm J$ models.

\medskip

{\bf Theorem 2.\/}\cite{NNS}  For any $d$ and any $\mu$ with finite mean,
(with $P$-probability one) at each 
site $x$ in $\Z^d$, there are only finitely many
flips with $\Delta H_x (\sigma) < 0$. 

\medskip

The cases left open by the results of \cite{NNS} were: (i) the homogeneous
ferromagnet or antiferromagnet for $d \ge 3$, (ii) $\pm J$ models for $d \ge 2$,
(iii) other noncontinuous $\mu$'s for $d\ge2$ and finally (iv) general continuous
$\mu$'s with infinite means for $d\ge2$. The main results of this paper are 
the following two theorems that resolve (ii) for $d=2$ and (iii) for $d\ge2$.

We remark that part of the proof of Theorem 4 can be easily applied to show
that $\rho_\F > 0$ for {\it any\/} continuous $\mu$; thus the $\mu$'s of
(iv) must either be type $\F$ or type $\M$.  Our guess is that (iv) is type
$\F$ for any $d\ge2$. As for (i), there is some numerical evidence
\cite{St} that homogeneous models remain type $\I$ for $d=3$ but perhaps
not for $d>4$. For more discussion of physical background and open
problems, see \cite{NS99,NS99b,NS00}.

\medskip

{\bf Theorem 3.\/}  $\pm J$ models are type $\M$ for $d=2$.

\medskip

{\bf Theorem 4.\/}  For any $d\ge2$, if $\mu$ is neither continuous nor of the form 
$\a \d_J + (1-\a) \d_{-J}$ for some $J\ge0$ and $0 \le \a \le 1$, then
$\mu$ is type $\M$.

\medskip

The proof of Theorem 4, presented in Section 2 of the paper, is quite easy.
In Sections 3 and 4, we give the proof of Theorem 3; the demonstration that 
$\rho_\I > 0$ (Section 3) is fairly easy but 
the proof that $\rho_\F  > 0$ (Section 4) is not.
The arguments used for the latter may be of general interest.
In the proofs of both parts of Theorem 3, an important role is played
by the frustration/contour representation of the $\pm J$ model for $d=2$
(see, e.g.,
\cite{Toulouse,FHS,BMRU,Barahona,BF}); there are natural extensions
of this representation for $d \ge 3$ that could
be useful in determining the type in these higher dimensions. 

As we shall see, there is an interesting conceptual difference between the 
proofs of these two theorems. The proof of Theorem 4 is essentially
{\it local\/} in that we demonstrate that certain sites are type $\I$
and certain are type $\F$ from knowledge of the couplings and spins
(at time zero) in finite regions containing those sites. 
The proof of Theorem 3, on the other
hand, is not local, in that using the local knowledge, we only manage
to deduce that {\it some\/} site among a finite number must be type $\I$
(or $\F$). This is because we are unable to find local configurations
of couplings and spins that completely insulate a local region 
from the surroundings. The (unknown in advance) influence
from the outside prevents a determination of the type of individual sites. 
Although we have not proved it, we suspect that $\pm J$ models are
intrinsically {\it nonlocal\/} in the strong sense that the type of any
site {\it cannot\/} be ascertained from strictly local knowledge.

\section{Other than $\pm J$ Models: Proof of Theorem 4}
\label{sec:otherproof}

{\bf Proof that $\rho_\F > 0$.\/}  Let $C$ denote some cube in $\Z^d$, such as the
unit cube consisting of vertices $x=(x_1,\dots,x_d)$ with each $x_i=0$ or $1$, and
let ${\cal E}_i(C)$ (resp., ${\cal E}_o(C)$) denote the set of nearest-neighbor
edges $\{x,y\}$ such that both $x$ and $y$ (resp., exactly one of $x$ and $y$)
belong to $C$. It suffices to show that with positive probability, the
$J_{x,y}$'s for $\{x,y\} \in {\cal E}_i(C) \cup {\cal E}_o(C)$ and the 
$\s_x^0$'s for $x \in C$ are such that $\s_x^t$ never flips for $x \in C$.

To do this, we note that for each $\mu$ of Theorem 4, $|J_{x,y}|$ is nonconstant.
Hence there exists some $J' > 0$ such that
$P(A_{J'}^+ (C) \cup (A_{J'}^- (C)) > 0$, where $A_{J'}^{\pm} (C)$ is the event
that $|J_{x,y}| > J'$ and sgn($J_{x,y}$) is the constant value $\pm 1$ for every
$\{x,y\}$ in ${\cal E}_i(C)$, while for every $\{x,y\}$ in ${\cal E}_o(C)$,
$|J_{x,y}| \le J'$. Let $B^+(C)$ (resp., $B^-(C)$) denote the event that
$(\s_x^0:\, x \in C)$ is constant (resp., is one of the two checkerboard
patterns). Then either $A_{J'}^+ (C) \cap B^+(C)$ or $A_{J'}^- (C) \cap B^-(C)$
(or both) have positive probability. But the occurrence of either one implies that
$\s_x^t$ never flips for $x$ in $C$ and so $\rho_\F > 0$.

To show that $\rho_\I > 0$, we use a slightly more complicated geometric
construction involving a site $w$ (e.g., the origin) and $2d$ disjoint
cubes $C_1,\dots,C_{2d}$ that are neighbors of $w$ in the sense that for
each $j$, $w \notin C_j$ but $ C_j$ contains exactly one nearest neighbor
$z_j$ of $w$ (see Fig.~1).  Since $\mu$ is not continuous, it has an atom
at some value ${\tilde J}$. We again construct events involving the
couplings and spins near $w$, but now the construction depends on which of
two cases $\mu$ falls into.

\begin{figure}
\centerline{\epsfig{file=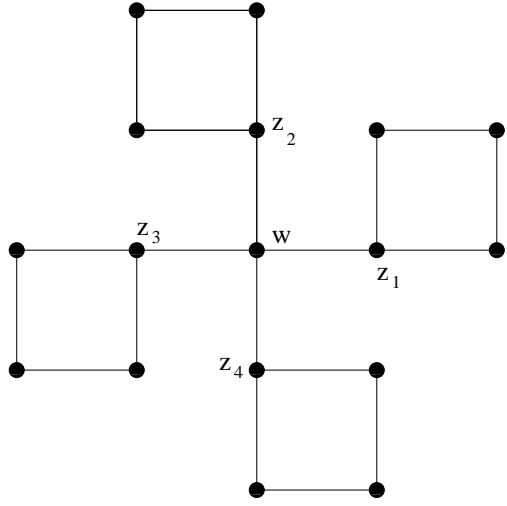,width=2.6in}}
\vspace{0.25in}
\caption{
Geometric construction demonstrating that a positive fraction of
spins flip infinitely often for all $d\ge 2$, in noncontinuous disordered
systems other than $\pm J$ models.  In this $d=2$ figure, filled circles
and solid lines denote respectively sites and edges of the original $\Z^2$
lattice.  Here, the spin at site $w$ flips infinitely often, given the
events discussed in the text in Sec.~2.}
\label{fig:propellor}
\end{figure}

{\bf Proof that $\rho_\I > 0$; Case 1.\/}  Suppose $\mu = \a \d_J + \beta \d_{-J} +\nu$
with $J > 0$, $0< \a + \beta <1$ and a continuous $\nu$ supported on
$[-J,J]$. Then either $J$ or $-J$ (or both if $\a,\beta > 0$) will work for
${\tilde J}$. We now define $D_{{\tilde J},j}$ to be the event that
$J_{x,y} = {\tilde J}$ for every $\{x,y\}$ in ${\cal E}_i(C_j)$
and for $\{x,y\} = \{z_j,w\}$, but for every other $\{x,y\}$ in ${\cal E}_o(C_j)$,
$|J_{x,y}| < |{\tilde J}|$. 
The event $D_{\tilde J} \equiv \cap_{j=1}^{2d} D_{{\tilde J},j}$
has positive probability. Now, for either value of sgn(${\tilde J}$), consider
the events $B^+$ and $B^-$, defined as 
\begin{equation}
\label{eq:seven}
B^{\pm}\ = \ \cap_{j=1}^{2d}(B^{\pm}(C_j) \cap \{\s_{z_j}^0 = (-1)^j\})
\end{equation}
and note that $P(D_{\tilde J} \cap B^{{\rm sgn} ({\tilde J})}) > 0$. 
We claim that if
$D_{\tilde J} \cap B^{{\rm sgn} ({\tilde J})}$ occurs, 
then $\s_w^t$ flips infinitely many times and
thus $\rho_\I > 0$. To see this, note that, very much as in the proof above that
$\rho_\F > 0$, if $D_{{\tilde J},j} \cap B^{{\rm sgn}({\tilde J})}(C_j)$ occurs
(and here we use the fact that $d \neq 1$),
then no site in $C_j$ ever flips. If in addition $\s_{z_j}^0 = (-1)^j$ for
each $j$, then $w$ has at all times exactly $d$ neighbors with $\s_x = +1$
and $d$ with $\s_x = -1$, so its rate for flipping is always $1/2$ and
it will flip infinitely many times.

{\bf Proof that $\rho_\I > 0$; Case 2.\/}  For any $\mu$ satisfying the hypotheses of
Theorem 4 that is not in Case 1, ${\tilde J}$ may be chosen so that
$|J_{x,y}| > |{\tilde J}|$ with positive probability. We now define
$D_{{\tilde J},j}^+$ and $D_{{\tilde J},j}^-$ as
\begin{equation}
\label{eq:eight}
D_{{\tilde J},j}^{\pm}\ = \ A_{|{\tilde J}|}^{\pm} (C_j) \cap \{J_{z_j,w} = {\tilde J}\}
\end{equation}
and $D_{\tilde J}^{\pm} = \cap_{j=1}^{2d}D_{{\tilde J},j}^{\pm}$ and
note that $P(D_{\tilde J}^+ \cup D_{\tilde J}^-) > 0$.
With $B^{\pm}$ defined in (\ref{eq:seven}), we have that either
$D_{\tilde J}^+ \cap B^+$ or $D_{\tilde J}^- \cap B^-$ (or both)
have positive probability. But if either occurs, then, as in Case 1,
$\s_{z_j}^t = (-1)^j$ for all $t$ and $\s_w^t$ will flip infinitely many times,
which completes the proof.

\section{Two-Dimensional $\pm J$ Models: $\rho_\I>0$.}
\label{sec:pmjrhoi}

We begin this section by introducing the frustration/contour representation
for the $\pm J$ model that we will use throughout this section and the
next for the proof of Theorem 3. We then give the proof that $\rho_\I>0$,
which concludes with a general lemma about recurrence that will also be used
(many times) in the next section for the proof that $\rho_\F>0$.

The frustration/contour representation (see, e.g.,
\cite{Toulouse,FHS,BMRU,Barahona,BF}) uses variables $(\P,\G)$ associated
with the dual lattice $\Z^{2*} \equiv \Z^2 +(1/2,1/2)$, that are determined
by $(\J,\s)$.  A (dual) site in $\Z^{2*}$ may be identified with the
plaquette $p$ in $\Z^2$ of which it is the center, and is called {\it
frustrated\/} for a given $\J$ if an odd number of the four couplings
$J_{x,y}$ making up the edges of that $p$ are antiferromagnetic; $\P$ is
then the set of frustrated (dual) sites. Thus $\P$ is determined completely
by $\J$, and it is not hard to see that {\it every\/} subset $\P$ of
$\Z^{2*}$ arises from some $\J$. The edge $\{x,y\}^*$ in $\Z^{2*}$, dual to
(i.e., the perpendicular bisector of) the edge $\{x,y\}$ of $\Z^2$, is said
to be {\it unsatisfied\/} for a given $\J$ and $\s$, if ${\mathrm
sgn}(J_{x,y} \s_x \s_y) = -1$ (and satisfied otherwise); $\G$ is then the
set of unsatisfied (dual) edges. We say that $(\J,\s)$ {\it gives rise} to
$(\P,\G)$, and that $\G$ is compatible with $\J$ or with $\P$ if there
exists some $\s$ such that $(\J,\s)$ gives rise to $(\P,\G)$. We define
$\partial \G$, the boundary of $\G$, as the set of (dual) sites that touch
an odd number of (dual) edges of $\G$; then $\G$ is compatible with $\P$ if
and only if $\P = \partial \G$. A (site self-avoiding) path in $\Z^{2*}$
consisting of edges from $\G$
will be called a {\it domain wall\/}. For a given $\P$, domain walls can
terminate (i.e., with no possibility of continuation) only on frustrated
sites; this is because any termination site touches exactly one edge
of $\G$ and thus belongs to $\partial \G$ ($= \P$).

For a given $\J$ or $\P$, the Markov process $\s^t$ determines a 
process $\G^t$, that is easily seen to also be Markovian. The transition
associated with a spin flip at $x \in \Z^2$ is a local ``deformation'' 
of the contour $\G^t$ at the (dual) plaquette $x^*$ in $\Z^{2*}$ that contains
$x$ ; this deformation interchanges the satisfied and unsatisfied 
edges of $x^*$ and leaves the boundary $\P$ of $\G^t$ unchanged. The only
transitions with nonzero rates are those where the number of unsatisfied 
edges starts at $k=4$ or $3$ or $2$ and ends at $0$ or $1$ or $2$, 
respectively; 
transitions with $k=4$ or $3$
(resp., $k=2$) correspond to energy-lowering (resp., zero-energy) flips
and have rate $1$ (resp., $1/2$). We will continue to use the terms
flip, energy-lowering, etc.~for
the transitions of $\G^t$. 

\medskip

{\bf Proof that $\rho_\I>0$.}  This is by far the easier part of the proof
of Theorem 3 and uses a strategy that is only a slight extension of the 
type of argument used in the previous section to prove $\rho_\I>0$ in 
Theorem 4. As in that proof, we will consider an event of positive
probability, here denoted $D$, involving the frustration configuration
in a finite region (and thus the values of only finitely many couplings).
Unlike that proof, we will not then intersect $D$ with some event involving
$\s^0$ to insure that for all $t$, some site $x$ has a positive flip
rate. Instead, we will show that given $D$, and {\it any\/} spin (or contour)
configuration in a certain fixed square $C$ of $\Z^2$, 
there must be at least one site
in $C$ with a positive flip rate. This will insure that, conditional on $D$,
at least one site in $C$ will flip infinitely many times. 

The region $C$ is a $6 \times 6$ square of $\Z^2$ and $D$ is defined
in terms of $\P$ restricted to the $5 \times 5$ square $\L^*$ of 
sites of $\Z^{2*}$ contained within $C$. We choose $D$ as the event that the
frustrated sites of $\L^*$ are exactly the nine sites (out of $25$)
indicated in Figure 2. These nine sites consist of a center site
$w_c$ and four adjacent pairs of sites to the Southeast, Northeast,
Northwest and Southwest of $w_c$. 

\begin{figure}
\centerline{\epsfig{file=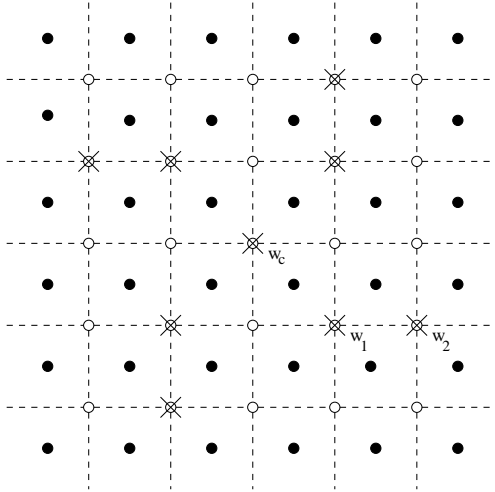,width=2.6in}}
\vspace{0.25in}
\caption{Geometric construction demonstrating that a positive fraction of
spins flip infinitely often in $\pm J$ models for $d=2$.  In this figure
the filled circles are sites in $\Z^2$, the empty circles are unfrustrated
sites in the (dual) $\Z^{2*}$ lattice, and each empty circle covered by an
$\times$ is a frustrated site in $\Z^{2*}$.  Dashed lines correspond to
edges in $\Z^{2*}$.  The significance of the $\Z^{2*}$-sites $w_c$, $w_1$,
and $w_2$ is discussed in Sec.~3.}
\label{fig:infinite}
\end{figure}

We have to show that for any
$\G$ compatible with $D$, 
there is at least one site in $C$ (or equivalently, one (dual)
plaquette touching $\L^*$) with a positive flip rate, i.e., with
at least two unsatisfied edges. 
In fact, if $D$ occurs, there must be a domain wall $\gamma_c$ starting from 
$w_c$; this is because $w_c$ is frustrated and so either one or three
of the edges touching it belong to $\G$. Either $\gamma_c$ has a ``bend''
within $\L^*$ and thus the (dual) plaquette just inside the bend has
a positive flip rate (since it has two or more unsatisfied edges)
or else $\gamma_c$ runs straight out of $\L^*$. In the latter case, by the invariance
of $D$ with respect to rotations by $\pi /2$, we may assume (without loss of 
generality) that $\gamma_c$ runs from $w_c$ to the East and passes just
above the (dual) edge joining the two Southeastern sites (that we will denote
$w_1, w_2$). But then there must be another domain wall $\gamma_1$
starting from $w_1$. Either $\gamma_c$ and $\gamma_1$ together determine
a positive flip rate site or else $\gamma_1$ runs from $w_1$
straight out of $\L^*$ to the South. But then there must be another domain
wall $\gamma_2$ starting from $w_2$, that (together with $\gamma_c$ and 
$\gamma_1$) will determine a positive flip rate site, no matter what
direction it runs off to. 

Let $A$ denote the set of $\G$ configurations such that
there is a site in $C$ with a
positive flip rate and let $B$ denote the event that there is a spin flip
in $C$ at some time $t \in [0,1]$. It is easy to see that 
for some $\alpha > 0$,
\begin{equation}
\label{eq:three-one}
\G \in A \, \Longrightarrow \, P(B|\G^0=\G) \ge \alpha. 
\end{equation}
It follows from Lemma 5 below
that conditional on $D$,
there will (with conditional
probability one) be infinitely many spin flips in $C$ and hence 
some site
in $C$ will flip infinitely many times. Since a positive density of the
translates of the event $D$ must occur (with probability one), we
conclude that $\rho_\I>0$ as desired.

\medskip

{\bf Lemma 5.\/}  Let $Z_t$ be a continuous-time Markov process with state 
space ${\cal Z}$ and time-homogeneous transition probabilities, 
and let $Z_t^{(\tau)}$ denote the time-shifted process
$Z_{\tau + t}$. For $A$ a (measurable) subset of ${\cal Z}$, say $A$
{\it recurs\/} if $\{\tau > 0:\, Z_\tau \in A \}$ is unbounded. For $B$
an event measurable with respect to $\{Z_t:\, 0 \le t \le 1\}$, say $B$
recurs if $\{\tau > 0:\, Z^{(\tau)} \in B\}$ is unbounded. If
\begin{equation}
\label{eq:three-two}
\inf_{z \in A}P(B|Z_0 = z) \ge \alpha >0 
\end{equation}
and $A$ recurs with probability one (resp., with positive
probability), then so does $B$.

\medskip

{\bf Proof.}  If $A$ occurs, then define 
$T_j$ inductively by $T_0 = 0$ and $T_{j+1}$ is the smallest
$\tau \ge T_j +1$ such that $Z_\tau \in A$. For $j \ge 1$,
let $\eta_j$ denote the indicator of the event that
$Z^{(T_j)} \in B$. It follows from (\ref{eq:three-two}), by conditioning
on the values of the $Z_{T_j}$'s, that $(\eta_1,\eta_2,\dots)$ stochastically
dominates $(\eta'_1,\eta'_2,\dots)$, a sequence of i.~i.~d. zero-or-one
valued random variables with $P(\eta'_j = 1) = \alpha$. Thus,
with probability one (conditional on $A$), 
$\sum \eta_j = \infty$ and $B$ recurs.
%\subsection{The Easy Part: $\rho_\I>0$.}
%\label{subsec:rhoi}
%
%\subsection{The Hard Part: $\rho_\F>0$.}
%\label{subsec:rhof}

\section{Two-Dimensional $\pm J$ Models: $\rho_\F>0$.} 
\label{sec:pmjrhof}

The general strategy in this section is somewhat
similar to that of the last section,
but the analysis is considerably more involved. We will again consider
an event, now denoted $D'$, involving the frustration configuration in a 
finite region $\L'$ of $Z^{2*}$, 
and the spin configuration in a fixed square $C'$
of $\Z^2$. Our object will be to show that at least one of the sites
in $C'$ will {\it eventually\/} have flip rate zero and hence will
flip only finitely many times, thus proving $\rho_\F>0$; this will be done
by proving that the domain wall geometry in $\L'$ must eventually satisfy
various constraints. The key
technique of the proof will be to combine Theorem 2 and
Lemma 5 to show that certain contour events $A$ are
{\it eventually absent (e-absent)\/}, \ie that 
$A$ recurs with probability {\it zero\/},
%$\{\tau > 0:\, \G^{(\tau)} \in A\}$
%is {\it bounded\/} with probability one,
since otherwise there would be infinitely many energy-lowering flips
in $\L'$ with positive probability. 

The region $\L'$ is an $8 \times 8$ square in $Z^{2*}$ and the event
$D'$ is that out of the $64$ sites in $\L'$, the frustrated ones are
exactly the $20$ sites indicated in Figure 3. These are all within
the ``border'' of $\L'$ and are those sites in the border that are at most 
distance two from one of the four corner sites. The region $C'$ 
is the $7 \times 7$ square $\C (\L')$
of sites of $\Z^2$ contained within $\L'$; 
these sites correspond to the $49$ dual plaquettes formed by the 
edges of $\L'$. As indicated in Figure 3, let $u_N$, $u_E$, $u_W$ and $u_S$ 
denote the sites in the exact middles of the North, East, South 
and West sides of the border of $C'$. 
What our proof will show is that eventually either $u_N$
(and $u_S$) or $u_E$ (and $u_W$) will have flip rate zero. The bulk of
the proof is a lengthy series of lemmas, most of which show 
that certain types
of contour configurations (in $\L'$) are e-absent.
\vfill\eject 

\begin{figure}
\centerline{\epsfig{file=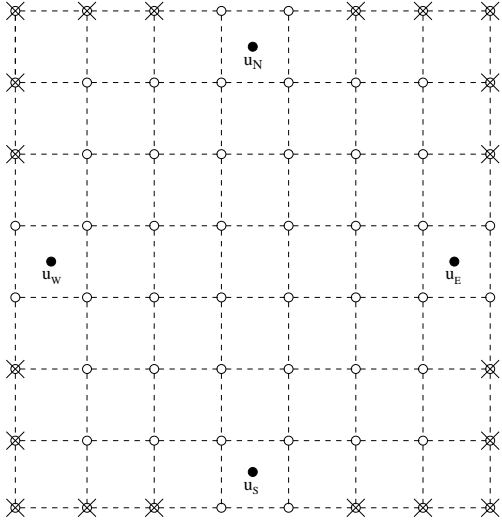,width=2.6in}}
\vspace{0.25in}
\caption{ Geometric construction demonstrating that a positive fraction of
spins flip only finitely many times in $\pm J$ models for $d=2$, as
explained in Sec.~4.  The conventions used in this figure are the same as
in Fig.~2.}
\label{fig:finite}
\end{figure}

Here is a sketch of how the lemmas
will lead to the desired conclusion. A contour configuration
$\G$ will be called of {\it horizontal\/} (resp., {\it vertical\/})
type if it contains a horizontal (resp., vertical)
domain wall, \ie one connecting the West and East (resp.,
South and North) sides of the border of $\L'$; a $\G$ that is of neither of
these two types will be said to be of {\it non-crossing\/} type. It turns out
(see Lemmas 12 and 13 below) that (conditional on $D'$) eventually
$\G^t$ is exclusively one of these three types --- \ie it will not 
simultaneously contain both a horizontal and a vertical domain wall, and
there will be no transition in which the type changes.
It also turns out (as a consequence of other lemmas and again conditional
on $D'$) that for $u_N$
(resp., $u_E$) to flip, $\G^t$ just before or just after the flip must
either be vertical (resp., horizontal) or else must be e-absent. It
follows that eventually at most one of $u_N$ and $u_E$ has positive flip
rate, completing the proof. Now to the lemmas.

In the lemmas, we will consider various rectangles, denoted $\L^*$ (or
sometimes $\Sigma^*$) of sites in $\Z^{2*}$, the associated rectangles
$\C(\L^*)$ of $\Z^2$-sites within $\L^*$, contour configurations $\G(\L^*)$
(and frustration configurations $\P(\L^*)$) restricted to $\L^*$, and {\it
internal\/} (or more specifically, $\L^*$-{\it internal\/}) transitions or
flips of these restricted contour configurations, \ie those corresponding
to (energy decreasing or zero-energy) flips of sites in $\C(\L^*)$ (these
do not include {\it external\/} flips, \ie of sites not in $\C(\L^*)$ that
are nearest neighbors of sites in $\C(\L^*)$). We will call $\G(\L^*)$ {\it
unstable\/} if it is the starting configuration of an energy decreasing
internal transition; \ie if $\G(\L^*)$ contains $3$ or $4$ edges of some
(dual) plaquette completely within $\L^*$.

\medskip

{\bf Lemma 6.\/}  Any unstable $\G(\L^*)$ is e-absent.

\medskip

{\bf Proof.\/}  This is an easy  consequence
of Theorem 2 and Lemma 5. Here $B$ is the event that an energy decreasing
internal flip takes place in a unit time interval and $\alpha$
may be bounded below by the probability that such a flip takes
place before any other (internal or external) flip that could change
$\G(\L^*)$. We leave further details to the reader.

\medskip

{\bf Lemma 7.\/}  Given compatible $\P(\L^*)$ and $\G(\L^*)$, if there
exists a rectangle $\S^* \supseteq \L^*$ such that for every $\G(\S^*)$
that coincides with $\G(\L^*)$ in $\L^*$ and is compatible with $\P(\L^*)$,
there is a finite sequence of $\S^*$-internal transitions,
$\G_1(\S^*) = \G(\S^*) \to \G_2(\S^*) \to \dots \to \G_n(\S^*)$,
(possibly with $n=1$) such that $\G_n(\S^*)$ is e-absent, then 
(conditional on $\P(\L^*)$) $\G(\L^*)$ is also e-absent. 

\medskip

{\bf Proof.\/}  For each $\G(\S^*)$ and each
$\S^*$-internal transition from that configuration, 
let $c_1(\Delta) > 0$ denote (a lower bound
for) the probability
that that transition 
is the first ($\S^*$-internal or $\S^*$-external)
flip to be attempted during a time interval of length $\Delta > 0$
and that flip is successful. Inductively, we see that with probability
at least $c(\G(\S^*)) = c_1(\frac{1}{n-1}) \cdots c_{n-1}(\frac{1}{n-1})$ 
(or $c(\G(\S^*)) = 1$ when 
$n=1$) $\G(\S^*)$ will transform into $\G_n(\S^*)$ sometime during a
time interval of unit length. We can now apply Lemma 5 with $B$ being the
event that one of these (finitely many) $\G_n(\S^*)$'s occurs during the
unit time interval and with
$\alpha$ being the minimum of the $c(\G(\S^*))$'s.

\medskip

A path or domain wall in $\Z^{2*}$ with endpoints $z$ and 
$w$ is called {\it monotonic\/} if, for one of the two directed
versions of the path, either every step 
moves to the East or to the North or else every step moves to the
East or to the South. For such a monotonic path $\gamma$, we denote by
$R(\gamma) = R(z,w)$ the (smallest) rectangle in $\Z^{2*}$
with $z$ and $w$ as two of its corners. 
For a non-monotonic $\gamma$, $R(\gamma)$ denotes the smallest rectangle
containing the sites of $\gamma$.

\medskip

{\bf Lemma 8.\/}  $\G(\L^*)$ is e-absent if it contains a non-monotonic
domain wall.

\medskip

{\bf Proof.\/} Any non-monotonic domain wall contains as a sub-path a
non-monotonic domain wall $\gamma$, with $R(\gamma)$ a $2 \times (m+1)$
rectangle and $\gamma$ going around one long and two short sides of the
border of the rectangle. Let $x_1,\dots,x_m$ denote the $\Z^2$-sites at the
centers of the $m$ (dual) plaquettes of $R(\gamma)$ (listed in either of
the two natural orders).  Consider the sequence of flips of the first $m-1$
of these sites (in the same order). If $\G(\L^*)$ contains no other edges
of $R(\gamma)$ than those of $\gamma$, then that sequence of flips
corresponds to a sequence of transitions as in Lemma 7 (with $\S^* = \L^*$)
whose final configuration is unstable; if there are other edges, then an
unstable configuration may be reached earlier. In either case, we conclude
from Lemmas 6 and 7 that $\G(\L^*)$ is e-absent.

\medskip

{\bf Lemma 9.\/}  Let $\G$ contain a monotonic domain wall $\gamma$
(with endpoints $z$ and $w$) but no other edge inside the rectangle 
$R(\gamma)$. If $\gamma'$ is any other monotonic path between $z$ and 
$w$, then there is a finite sequence of $R(\gamma)$-internal
flips (i.e., flips of $\Z^2$-sites within
$R(\gamma)$) that transforms $\G$ into a configuration $\G'$ whose restriction
to $R(\gamma)$ consists exactly of the edges of $\gamma'$. 

\medskip

{\bf Proof.\/} We sketch a proof, but the reader is invited to provide her
own for this elementary result.  Suppose (without loss of generality) that
$z$ is the Southwest and $w$ the Notheast corner of $R(\gamma)$. Let
$\gamma''$ denote the path between those corners that runs along the South
and East sides of the rectangle. It suffices to show that any $\gamma$ (and
hence also $\gamma'$) can be transformed into $\gamma''$ (and vice-versa,
by inversion). But this can be done (inductively) by noting that for any
$\gamma \neq \gamma''$, there is some site within $R(\gamma)$ whose flip
will strictly reduce the area of the region between $\gamma$ and
$\gamma''$.

\medskip

In an $m \times n$ rectangle $\L^*$ of $\Z^{2*}$, the border consists of
those sites in $\L^*$ that are nearest neighbors of sites outside $\L^*$.
The border has four (distinct, unless $m=n = 1$, but not disjoint) sides:
North, East, West and South. There are four corners (distinct, if $m,n > 1$),
denoted NE, NW, SW and SE, each of which is the single site at the intersection
of two adjacent sides of the border. We define the {\it interior\/}
of $\L^*$ (denoted int$(\L^*)$) as those sites in $\L^*$, that are not in its
border and we define the interior of any side of the border as those sites
in that side that are not corners. 

\medskip

{\bf Lemma 10.\/} Given an $m \times n$ rectangle $\L^*$ 
with $m,n>1$ and conditional on 
a frustration configuration
$\P(\L^*)$, a contour configuration $\G(\L^*)$ is e-absent if
it contains a monotonic domain wall $\gamma$ between some $z$ and $w$
and any one of the following
four situations holds for the rectangle $R = R(\gamma) = R(z,w)$:

(i) $\G(\L^*)$ contains an edge $e^*$ in $R$ that is not in $\gamma$.

(ii) There is a frustrated site in int($R$).

(iii) There are two frustrated sites in the 
interior of a single side of the border of $R$. 

(iv)  A corner of $R$ other than $z,w$ is frustrated and so is at least
one site in the interior of each of the two sides of the border
of $R$ touching that corner. 

\medskip

{\bf Proof.\/}  (i) Let $\gamma'$ be any monotonic path between
$z$ and $w$ that contains
$e^*$.  Consider the sequence of flips provided by Lemma 9 that would
(if there were no edges of $\G(\L^*)$ in $R$ other than those of
$\gamma$) transform $\gamma$ into $\gamma'$. Because $e^*$ is 
already in $\G(\L^*)$ (and so may be
other edges of $R$ that are not in $\gamma$),
at some stage along this sequence of flips (before $e^*$ is absorbed into
the evolving domain wall) $\G(\L^*)$ will have been transformed into
an unstable configuration. The desired conclusion then follows from Lemmas 
6 and 7. 

(ii) Since a frustrated site must have an odd number of unsatisfied edges
touching it, such a site in the interior of $R$ has an unsatisfied
edge $e^*$ not in $\gamma$ (but in $R$) touching it. The result now follows
from part (i). 

(iii) Without loss of generality, we assume that $z$ and $w$ are the SW and
NE corners of $R$ and the two frustrated sites are on the south side of the
border of $R$. Since these are not endpoints of $\gamma$, but they are
frustrated, they must each have at least one unsatisfied edge not from
$\gamma$ touching them. By part (i), we may assume that those edges go out
from $R$ to the South. Also, by part (i) we may assume that $\G(\L^*)$ has
no edges other than those of $\gamma$ in $R$. Then by the sequence of flips
provided by Lemma 9, $\G$ (whose restriction to $\L^*$ is $\G(\L^*)$) can
be transformed into $\G''$ where $\gamma$ is replaced by $\gamma''$, a
domain wall between $z$ and $w$ lying along the South and East sides of
$R$. But $\G''$ also contains the two edges going South from the two
frustrated sites. Thus it contains a non-monotonic domain wall in the
slightly larger region $\S^*$, that adds to $\L^*$ its neighboring
sites. The desired conclusion now follows from Lemmas 7 and 8.

(iv) We may assume that $z$ and $w$
are the SW and NE corners of $R$, that there is a frustrated site 
in the interior of each of the South and East sides of the border and that the
SE corner is also frustrated. By the same reasoning as in part (iii),
there must be an unsatisfied edge going out from $R$
starting from each of these three frustrated
sites. The ones from the interior sites on the sides go to the South and
the East, while the one from the corner can go in either of those two 
directions. Thus there are either two unsatisfied edges going South
from the South side or else two going East from the East side. In either
case, the proof is completed as in part (iii). 

\medskip

We now focus on the $8 \times 8$ square $\L'$ and the frustration
configuration $\P'(\L')$ (or the event $D'$ that the frustration
configuration in $\L'$ is exactly $\P'(\L'))$ indicated in Figure 3. Since
there are no frustrated sites in int($\L'$), domain walls of any $\G(\L')$
compatible with $\P'(\L')$ must be extendable so that the endpoints $z$ and
$w$ are both on the border of $\L'$. By Lemma 8, if $z$ and $w$ are on a
single side of the border and $\G(\L')$ is not e-absent, then the domain
wall can only be the straight line path between $z$ and $w$.  The following
two lemmas cover the situations where the endpoints are on adjacent or
opposite sides and give restrictions on the possible $\G(\L')$'s that are
not e-absent. In the first of the two lemmas we write $|z-z'|$ to denote
the Euclidean distance between sites in $\Z^{2*}$.

\medskip

{\bf Lemma 11.\/}  Condition on $D'$. Every $\G(\L')$ that
contains a domain wall between sites $z$ and $w$ that are on adjacent
sides (but not on any single side) 
of the border of $\L'$,
is e-absent if $z,w$ and the common corner $c(z,w)$ of the two sides 
do not satisfy the following condition:
\begin{equation}
\label{eq:four-one}
|z-c(z,w)| + |w - c(z,w)| \le 3.
\end{equation}
(If $z$ and $w$ are opposite corners of $\L'$, then $c(z,w)$ can be taken
as either of the two remaining corners, the condition is not satisfied and
$\G(\L')$ is e-absent.)  Every $\G(\L')$ that contains a domain wall
$\gamma$ between $z$ and $w$ on a single side of the border of $\L'$ with
$z$ a corner, is e-absent unless
\begin{equation}
\label{eq:four-two}
|z-w| \le 2.
\end{equation}

\medskip

{\bf Proof.\/}  We may assume by Lemma 8
that the domain wall is monotonic.
If $z$ and $w$ are not on a single
side and (\ref{eq:four-one}) does not hold, then one of the
following two cases occurs. 

(I) One of $|z-c(z,w)|$ or $|w - c(z,w)|$ is at least $3$. In this case,
since we condition on $D'$, one of the sides of $R(z,w)$ contains at least
two frustrated sites in its interior and part (iii) of Lemma 10 applies.

(II) $|z-c(z,w)| = 2 = |w - c(z,w)|$. In this case, 
since we condition on $D'$, a corner (other than
$z$ or $w$) of $R(z,w)$ is frustrated and so is one site in 
the interior of each of the adjacent sides of the border of $R(z,w)$.
Thus, part (iv) of Lemma 10 applies.

If $z$ and $w$ are on a single side with $z$ a corner, we may assume,
without loss of generality, that $z$ is the NW corner and $w$ is on the
North side. If (\ref{eq:four-two}) does not hold, then there are (at least)
two frustrated sites on $\gamma$ between $z$ and $w$, and there must be
unsatisfied edges not in $\gamma$ touching these two sites.  One of those
edges must go to the South (into $\L'$), or else there would be a
non-monotonic domain wall (using some edges of $\gamma$ and two edges going
North just outside of $\L'$). Since there is no frustration in int($\L'$),
that South-going edge must be extendable to a domain wall reaching some
site $w'$ on the border of $\L'$. Combining that extension with part of
$\gamma$ yields a domain wall $\gamma'$ between $z$ and $w'$.  If $w'$ is
on the West or North sides, $\gamma'$ would be non-monotonic.  If $w'$ is
on the South or East sides, then (\ref{eq:four-one}) with $w$ replaced by
$w'$ would not hold and $\G(\L')$ would be e-absent by the part of this
lemma that has already been proved.

\medskip

Before stating the next lemma, we recall our definition of a horizontal
(resp., vertical) domain wall in $\L'$ as one whose endpoints are in
the West and East (resp., South and North) sides of the border. 

\medskip

{\bf Lemma 12.\/}  Conditional on $D'$, every $\G(\L')$ that contains
{\it both\/} a vertical and horizontal domain wall is e-absent.

\medskip

{\bf Proof.\/} Let us denote the endpoints of the 
horizontal (resp., vertical) domain
wall by $z_W$ and $z_E$ (resp., $z_S$ and $z_N$) so that the subscript
indicates the side that the endpoint is located on. (Note though
that the endpoints may
be corners.) Since the vertical and horizontal domain walls must have
at least one site of $\L'$ in common, it follows that $\G(\L')$ has a
domain wall with {\it any\/} pair of the points $\{z_N,z_E,z_W,z_S\}$ as
endpoints. 
%Let us denote by $c_{SW}, c_{SE},\dots$ the corners of
%$\L'$. Noting that $|z_S-c_{SW}|+|z_S-c_{SE}| = 7$, we see that
%without loss of generality, we may assume that $|z_S-c_{SE}| \ge 4$,
%and similarly that $|z_E-c_{SE}| \ge 4$. 

It also follows that both the horizontal and vertical crossings must
be straight lines or else there would be a non-nonotonic domain wall.
Hence $z_S, z_N$ are at distance $\ge 4$ from one of the East or
West sides (which we take to be the East side, without loss of generality)
and similarly (without loss of generality) $z_W, z_E$ may be assumed
to be at distance $\ge 4$ from the South side.
But then the domain wall with endpoints $z=z_S$ and $w=z_E$ violates
(\ref{eq:four-one}) and $\G(\L')$ is e-absent by Lemma 11. 

\medskip

We recall that $\G(\L')$ is said to be of horizontal or vertical or
non-crossing type according to whether it contains a horizontal or a vertical
domain wall or neither. We will also say that $\G^t$ is 
{\it eventually of type A\/} if for some (random) finite $T$, $\G^t$ is
of type A for all $t \ge T$, A standing for one of the above three types. 

\medskip

{\bf Lemma 13.\/} Suppose $\G(\L')$ and $\G'(\L')$ are related by an
internal or external flip (either $\G(\L')\to \G'(\L')$
or $\G'(\L') \to \G(\L')$), and suppose further that $\G(\L')$ has
a vertical (or horizontal) domain wall but $\G'(\L')$ does not. Conditional
on $D'$, any such $\G'(\L')$ is e-absent and, with probability one, $\G^t$ 
is eventually of one of the three types --- vertical, horizontal or non-crossing. 

\medskip

{\bf Proof.\/} Let $\gamma$ be a vertical domain wall in $\G(\L')$ between
$z$ on the South side and $w$ on the North side. The
flip changes the edges of a single (dual) plaquette in or next to $\L'$;
whether the flip is internal or external, zero-energy or energy lowering,
there will remain in $\G'(\L')$ a 
portion of $\gamma$ from $z$ to some $z' \in \L'$ on
that plaquette and a portion from $w$ to some $w' \in \L'$ on
that plaquette. Without loss of generality, we may assume that the distance
from $z'$ to the South side is at least $3$, and we then denote 
by $\gamma'$ the domain wall
portion from $z$ to $z'$. 

If $z'$ is a border site, it cannot be on the North side since then
$\gamma'$ would violate the assumption that $\G'(\L')$ is not of vertical
type.  For either the East or West side as a location for $z'$, it would
follow that $\gamma'$ is either non-monotonic or else $\G'(\L')$ is
e-absent because of violating (\ref{eq:four-one}) or (\ref{eq:four-two})
with $w$ replaced by $z'$.

If $z'$ is not a border site, then it is unfrustrated and $\gamma'$ must be
extendable to a $\gamma''$ between $z$ and some border site $z''$.  The
e-absence of $\G'(\L')$ now follows by the same arguments as above but with
$z'$ and $\gamma'$ replaced by $z''$ and $\gamma''$. Of course, analogous
arguments work when $\G(\L')$ is of horizontal rather than vertical type.

The final claim of the lemma now follows by choosing $T$ to be the
finite (with probability one) time beyond which no e-absent configurations
in $\L'$ are taken on by $\G^t$. By Lemma 12 and the part of this lemma already
proved, no changes of type occur after that time. 

\medskip

{\bf Proof that $\rho_{\F} > 0$.} Since $D'$ occurs with strictly
positive probability (and hence a positive density of translates of
$D'$ occur with probability one), it suffices to show that
conditional on $D'$, with probability one, for times
beyond some finite $T$, some $Z^2$-site in the $7 \times 7$ square
$\C(\L')$ will not flip. We take the same $T$ as in the proof of
the previous lemma, namely the time beyond which no e-absent
$\G(\L')$'s are seen. Past this time, $\G^t$ remains of one particular type,
and we will locate a non-flipping site depending on the type. 

Let $u_N$ denote the site in the middle of the North side of $\C(\L')$ (and
$u_E,u_W,u_S$ the sites in the middle of the other sides), as indicated in
Figure 3. A flip of $u_N$ corresponds to a change in $\G^t$ involving the
edges of the (dual) plaquette inside $\L'$ and just below the middle of its
North side. Since e-absent configurations are no longer seen, it must be a
zero-energy flip in which both before and after the flip there are exactly
two unsatisfied edges from that plaquette, but with the unsatisfied and
satisfied edges exchanged by the flip.  Thus either before or after the
flip, $\G^t$ must contain an edge going South from one of the two central
sites (that we will denote $z$) on the North side of $\L'$. That
South-going edge must be extendable to a domain wall $\gamma$ between $z$
and some other border site $w$. We claim that $\gamma$ must be vertical
because otherwise $\G^t$ would be e-absent. This is so because if $\gamma$
were not vertical, then $w$ would either be on the North side and so
$\gamma$ would be non-monotonic and e-absence would follow from Lemma 8; or
else $w$ would be on the West or East sides and e-absence would follow from
Lemma 11. This shows that after time $T$, $u_N$ (and by symmetry $u_S$)
cannot flip unless $\G^t$ is of vertical type.  Similarly $u_E$ and $u_W$
cannot flip after $T$ unless $\G^t$ is of horizontal type. By Lemma 13,
conditional on $D'$, after (the almost surely finite) time $T$ some site
(\eg either $u_N$ or $u_E$) does not flip. This completes the proof.

\medskip 

{\it Acknowledgments.\/} This research was supported in part by a Fulbright
grant and by the Italian MURST, cofin '99, under the Research Programme
``Stochastic Processes with Spatial Structure'' (AG), and by NSF Grants
DMS-98-02310 (CMN) and DMS-98-02153 (DLS).  A.~G. thanks Joel Lebowitz and
Rutgers University for their hospitality. A.~G. and C.~M.~N. thank Anton
Bovier and WIAS, Berlin for their hospitality.


\begin{references}

\small


\bibitem{NNS} 
S.~Nanda, C.M.~Newman, and D.L.~Stein, Dynamics of Ising Spin
Systems at Zero Temperature, pp. 183-194 in {\it On Dobrushin's Way (from
Probability Theory to Statistical Physics)\/}, R.~Minlos, S.~Shlosman and
Y.~Suhov, eds.,~Amer.~Math.~Soc.~Transl.~(2)~198~(2000).

\bibitem{NS99}
C.M.~Newman and D.L.~Stein, Blocking and Persistence in the Zero-Temperature
Dynamics of Homogeneous and Disordered Ising Models, 
Phys.~Rev.~Lett.~{\bf 82}, 3944--3947 (1999).

\bibitem{Jainold}
S.~Jain, Zero-Temperature Dynamics of the Weakly Disordered Ising Model,
Phys.~Rev.~E {\bf 59}, R2493--R2496 (1999).

\bibitem{Jainnew}
S.~Jain, Persistence in the Zero-Temperature Dynamics of the Diluted
Ising Ferromagnet in Two Dimensions, Phys.~Rev.~E {\bf 60}, R2445--R2447 (1999).

\bibitem{FIN}
L.R.G.~Fontes, M.~Isopi and C.M.~Newman, Chaotic Time Dependence in a 
Disordered Spin System, Prob.~Theory Rel.~Fields {\bf 115}, 417--443 (1999). 

\bibitem{Bray}
A.J.~Bray, Theory of Phase-Ordering Kinetics, Adv.~Phys.~{\bf 43}, 357--459 (1994). 

\bibitem{St}
D.~Stauffer, Ising Spinodal Decomposition at $T=0$ in One to Five
Dimensions, J.~Phys.~A {\bf 27}, 5029--5032 (1994).

\bibitem{De}
B.~Derrida, Exponents Appearing in the Zero-Temperature Dynamics of the
$1D$ Potts Model, J.~Phys.~A {\bf 28}, 1481--1491 (1995).

\bibitem{DHP}
B.~Derrida, V.~Hakim and V.~Pasquier, Exact First-Passage Exponents of
$1D$ Domain Growth:  Relation to a Reaction-Diffusion Model, 
Phys.~Rev.~Lett.~{\bf 75}, 751--754 (1995).

\bibitem{MH}
S.N.~Majumdar and D.~Huse, Growth of Long-Range Correlations after a Quench
in Phase-Ordering Systems, Phys.~Rev.~E {\bf 52}, 270--284 (1995).

\bibitem{MS}
S.N.~Majumdar and C.~Sire, Survival Probability of a Gaussian
Non-Markovian Process:  Application to the $T=0$ Dynamics of the
Ising Model, Phys.~Rev.~Lett.~{\bf 77}, 1420--1423 (1996).

\bibitem{CK}
J.~T.~Cox and A.~Klenke, Recurrence and Ergodicity of Interacting 
Particle Systems, Prob.~Theory Rel.~Fields {\bf 116}, 239--255 (2000).

\bibitem{NS99b}
C.~M.~Newman and D.~L.~Stein, Equilibrium Pure States and Nonequilibrium
Chaos, J.~Stat.~Phys.~{\bf 94}, 709--722 (1999).

\bibitem {EA} 
S.~Edwards and P.W.~Anderson, Theory of Spin Glasses, 
J.~Phys.~F {\bf 5}, 965--974 (1975).

\bibitem {BY}
K.~Binder and A.P.~Young, Spin Glasses: Experimental Facts,
Theoretical Concepts, and Open Questions, Rev.~Mod.~Phys.~{\bf 58}, 
801--976 (1986).

\bibitem{Arr}
R.~Arratia, Site Recurrence for Annihilating Random Walks on ${\bf Z}_d$,
Ann.~Prob.~{\bf 11}, 706--713 (1983).

\bibitem {CG}
J.T.~Cox and D.~Griffeath, Diffusive Clustering in the Two Dimensional
Voter Model, Ann.~Prob.~{\bf 14}, 347--370 (1986).

\bibitem{NN}
S.~Nanda and C.M.~Newman, Random Nearest Neighbor and Influence Graphs
on $\Z^d$, Ran.~Structures and Algorithms {\bf 15}, 262--278 (1999).

\bibitem{NS00}
C.~M.~Newman and D.~L.~Stein, Zero-Temperature Dynamics of Ising 
Spin-Systems Following a Deep Quench: Results and Open Problems, 
Physica A {\bf 279}, 159--168 (2000).

\bibitem{Toulouse}
G.~Toulouse, Theory of the Frustration Effect in Spin Glasses. I, 
Commun. Phys.~{\bf 2}, 115--119 (1977).

\bibitem{FHS}
E.~Fradkin, B.~Huberman, and S.H.~Shenker, Gauge Symmetries in Random
Magnetic Systems, Phys.~Rev.~B {\bf 18}, 4789--4814 (1978).

\bibitem{BMRU}
L.~Bieche, J.P.~Uhry, R.~Maynard, and R.~Rammal, 
On the Ground States of the Frustration Model of a Spin Glass by a Matching
Method of Graph Theory, J.~Phys.~A {\bf 13}, 2553--2576 (1980).

\bibitem{Barahona}
F.~Barahona, On the Computational Complexity of Ising Spin Glass Models,
J.~Phys.~A {\bf 15}, 3241--3253 (1982).

\bibitem{BF}
A.~Bovier and J.~Fr{\" o}hlich, A Heuristic Theory of the Spin Glass
Phase, J.~Stat.~Phys. {\bf 44}, 347--391 (1986).


\end{references}
\end{document}